\documentclass[onecolumn,prl,twocolumn,superscriptaddress]{revtex4}
\usepackage[latin9]{inputenc}
\usepackage{color}
\usepackage{amsmath}
\usepackage{amssymb}
\usepackage{graphicx}
\usepackage{tikz}
\usepackage{amsfonts}
\usepackage{graphicx} 
\usepackage{dsfont}
\usepackage{algorithm}
\usepackage{algorithmic}
\usepackage{multirow}

\usepackage{mathrsfs} 
\usepackage{float}
\usepackage{bbm}



\usepackage{epstopdf}
\usepackage{epsfig} 
\usepackage{verbatim}
\usepackage{array}
\usepackage{setspace}

   \allowdisplaybreaks[1]
   
\usepackage[unicode=true,
 bookmarks=true,bookmarksnumbered=false,bookmarksopen=false,
 breaklinks=false,pdfborder={0 0 1},backref=false,colorlinks=true]
 {hyperref}
\hypersetup{
 linkcolor=magenta, urlcolor=blue, citecolor=blue, pdfstartview={FitH}, hyperfootnotes=false, unicode=true}

\makeatletter
\@ifundefined{textcolor}{}
{%
 \definecolor{BLACK}{gray}{0}
 \definecolor{WHITE}{gray}{1}
 \definecolor{RED}{rgb}{1,0,0}
 \definecolor{GREEN}{rgb}{0,1,0}
 \definecolor{BLUE}{rgb}{0,0,1}
 \definecolor{CYAN}{cmyk}{1,0,0,0}
 \definecolor{MAGENTA}{cmyk}{0,1,0,0}
 \definecolor{YELLOW}{cmyk}{0,0,1,0}
}

\usepackage{amsfonts}\usepackage{tabularx}\usepackage{dcolumn}\usepackage{bm}\usepackage{graphicx}\usepackage{epstopdf}

\hypersetup{urlcolor=blue}

\makeatother

\newcolumntype{C}[1]{>{\centering\arraybackslash$}p{#1}<{$}}

\begin{document}

\title{Group Sparse Matrix Optimization for Efficient Quantum State Transformation}

\author{Kin Man Lai}
\affiliation{Department of Physics, City University of Hong Kong, Tat Chee Avenue, Kowloon, Hong Kong SAR, China}
\author{Xin Wang}
\email{x.wang@cityu.edu.hk}
\affiliation{Department of Physics, City University of Hong Kong, Tat Chee Avenue, Kowloon, Hong Kong SAR, China}
\affiliation{City University of Hong Kong Shenzhen Research Institute, Shenzhen, Guangdong 518057, China}
\date{\today}

\begin{abstract}
Finding ways to transform a quantum state to another is fundamental to quantum information processing. In this paper, we apply the sparse matrix approach to the quantum state transformation problem. In particular, we present a new approach for searching for unitary matrices for quantum state transformation by directly optimizing the objective problem using the Alternating Direction Method of Multipliers (ADMM).
Moreover, we consider the use of group sparsity as an alternative sparsity choice in quantum state transformation problems. Our approach incorporates sparsity constraints into quantum state transformation by formulating it as a non-convex problem. It establishes a useful framework for efficiently handling complex quantum systems and achieving precise state transformations.
\end{abstract}

\maketitle

\section{introduction} \label{sec:Intro}
Quantum information processing relies on the accurate and efficient manipulation of quantum states. Quantum state transformation aims to find specific transformations that can steer one quantum state to another. Specifically, given an initial state $|\psi_{i}\rangle$ and target state $|\psi_{t}\rangle$, quantum state transformation aims to search for an optimal matrix transformation $\mathbf{X}$ such that $\mathbf{X}|\psi_{i}\rangle$ and $|\psi_{t}\rangle$ are as close as possible to each other.

As the number of qubits increases, the dimension of the quantum state grows exponentially, making traditional methods tackling the problem computationally expensive. Recently, there arises interest to apply sparse optimization techniques to address the computational complexity of the problem. By exploiting the inherent sparsity that can arise from physical considerations or the characteristics of the quantum system, sparse optimization enables the identification of essential components while discarding less significant ones. Several studies have focused on generating unitary transformations for various quantum processing applications~\cite{seleznyova1995unitary, palao2002quantum, ho2009landscape, tibbetts2012exploring}.

Remarkable progresses have been made in applying the sparse matrix method to quantum state transformation and related problems. Refs.~\cite{gleinig2021efficient, de2022double} consider transforming sparse quantum states, which contain many zero coefficients, into desired states. Ref.~\cite{malvetti2021quantum} exploits the sparse isometry property of the state for quantum state transformation.
These methods take advantage of the sparsity inherent in the state by considering the sparse characteristics of the variables involved. 
Other applications include Ref.~\cite{yang2017effective}, which introduces sparsity on the noise while optimizing for a unitary density matrix for quantum tomography. 
Nevertheless, an efficient method to directly sparse the transformation matrix has been lacking in the literature.

On the other hand, group sparsity has emerged as a powerful concept in optimization, considering the interdependencies and natural groupings among system parameters
\cite{zhang2014group, liu2015image, pan2021group, gao2015multi, cho2014robust, zeng2018robust}.
This sparsity property allows subsets of matrix elements to be simultaneously zero or near-zero, leading to a significant reduction in computational complexity and improvement in efficiency. Specifically, a type of group sparsity known as row (or column) sparsity encourages solutions to exhibit sparsity patterns at the row (or column) level, allowing for advantageous in certain problems. Group sparsity finds applications in various fields, including computer vision, where it plays a crucial role in compressive sensing problems like image recovery~\cite{zhang2014group, liu2015image, pan2021group} and action recognition~\cite{gao2015multi, cho2014robust, zeng2018robust}, as well as medical sciences including neurobiology~\cite{zhang2018temporally, jiao2018sparse, beer2019incorporating}, pathology~\cite{wang2020block, huo2020sgl}, and genetic research~\cite{che2020genetic}. 
Ref.~\cite{maddu2021learning} aims to enforce conservation laws and guarantee symmetries when learning or inferring differential equation models. These studies highlight the potential of group sparsity in addressing challenges and improving results in optimization problems with a grouped structure in various domains, often outperforming regular sparsity~\cite{huang2010benefit}.
Employing group sparsity techniques in the field of quantum state transformation has the potential to enhance the efficiency and accuracy of transformation matrices that represent quantum circuits.

We propose a method to tackle the quantum state transformation problem using the Alternating Direction Method of Multipliers (ADMM) method. ADMM is an iterative optimization method that effectively handles problems with multiple constraints or variables by decomposing them into smaller sub-problems. This optimization method has shown success in various domains, including sparse optimization~\cite{chartrand2013nonconvex}, sparse networks~\cite{ma2023general}, and quantum state reconstruction~\cite{yang2017effective, li2016improved, li2017fast}. 
Our method is designed to identify the minimum values of convex objective function subject to unitary and equality constraints, while adhering sparsity constraints. 
The algorithm aims to approximate and identify the specific unitary operator that can achieve the desired state transformation. Moreover, our method is well-suited for problems that can be modeled as a convex function with (or without) unitary constraints and equality constraints, making it applicable to a wide range of problems. One such example is compressed sensing, which has applications in  quantum state tomography~\cite{yang2017effective, gross2010quantum, smith2013quantum, li2017fast}. Additionally, to the best of our knowledge, our method is the first to   tackle the challenges of sparse optimization under both unitary and equality constraints. This new combination of constraints sets our approach apart from existing methods and highlights its usefulness in efficient quantum state transformations.



The remainder of this paper is organized as follows: We begin by introducing some useful notation in Section~\ref{sec:notation}. Section~\ref{sec:formulation} presents the mathematical formulations of the problem in detail. Our ADMM optimization method for searching sparse (or group sparse) transformation matrix is presented in Section~\ref{sec:method}. Furthermore, Section~\ref{sec:construction} delves into the realm of Quantum Circuit Decomposition methods. In Section~\ref{sec:results}, we present our comparative analysis and numerical results. Finally, we conclude the paper in Section~\ref{sec:conclusion} with a summary of our contributions and future research directions.

\section{Notations} \label{sec:notation}
 In this paper, we use the following notations to represent different norms:

\begin{itemize}

  \item $\|\mathbf{X}\|_p$: This represents the $\ell_p$ norm of a matrix $\mathbf{X}$. It is defined as the $p$-th root of the sum of the absolute values of its components raised to the power of $p$, where $p\neq0$ is a positive real number. Mathematically, it can be expressed for an $m\times n$ matrix $\mathbf{X}$ with elements $x_{ij}$, $i=1\ldots m$ and $j=1\ldots n$ as: 
\begin{equation} \label{eq:pnorm}
  \|\mathbf{X}\|_p = \left(\sum_{i=1}^m\sum_{j=1}^n |x_{ij}|^p \right)^{\frac{1}{p}} \;\;.
\end{equation}
\item $\|\mathbf{X}\|_2$: Specifically, for $p=2$ the norm denotes the Euclidean norm, commonly used to measure the similarity between two matrices for fidelity calculations, given as:
\begin{equation} \label{eq:l2norm}
  \|\mathbf{X}\|_2 = \left(\sum_{i=1}^m\sum_{j=1}^n |x_{ij}|^2 \right)^{\frac{1}{2}} \;\;.
\end{equation}
To ensure differentiability and facilitate optimization,  many optimization problems involving the Euclidean norm is often modified by taking the square of Eq.~\eqref{eq:l2norm}:
\begin{equation} \label{eq:l2norm_square}
  \|\mathbf{X}\|_2^2 = \sum_{i=1}^m\sum_{j=1}^n |x_{ij}|^2 \;\;.
\end{equation}

\item $\|\mathbf{X}\|_0$: When $p=0$, the norm $\|\mathbf{X}\|_0$ represents the number of non-zero elements in the matrix $\mathbf{X}$. It counts the total number of entries in $\mathbf{X}$ that are not equal to zero.
  
\item $ \|\mathbf{X}\|_{p,q}$: This notation refers to the $\ell_{p,q}$ norm, where $p,q\neq0$ are positive real numbers. 
$\|\mathbf{X}\|_{p,q}$ is calculated as follows:
\begin{equation} \label{eq:pq_norm}
\|\mathbf{X}\|_{p,q} = \left(\sum_{i=1}^m \left(\sum_{j=1}^n |x_{ij} |^p \right) ^{\frac{q}{p}} \right) ^{\frac{1}{q}} \;\;.
\end{equation}

\item $ \|\mathbf{X}\|_{2,0}$: This is a specific norm represented in Ref.~\cite{pang2018efficient},  which counts the number of non-zero rows in the matrix $\mathbf{X}$, defined as follows:
\begin{equation} \label{eq:20_norm}
\|\mathbf{X}\|_{2,0}  = \sum_{i=1}^m  \left\|{\sum_{j=1}^n x^2_{ij}}\right\|_0 \;\;.
\end{equation}

\end{itemize}

These norms are essential in shaping our objective function and establishing sparsity constraints in the following sections. Specifically, we consider general sparsity $\ell_{1}$ norm and row sparsity $\ell_{2,1}$ norm, defined as:
\begin{equation} \label{eq:l1norm}
\|\mathbf{X}\|_1 = \sum_{i=1}^m \sum_{j=1}^n |x_{ij}| \qquad (\ell_{1}\mathrm{-norm})\;\;,
\end{equation}
\begin{equation} \label{eq:l21norm}
\|\mathbf{X}\|_{2,1}  = \sum_{i=1}^m  \sqrt{\sum_{j=1}^n x^2_{ij}} \qquad (\ell_{2,1}\mathrm{-norm})\;\;.
\end{equation}

Additionally, For ease of reference, we denote identity matrix as $\mathbf{I}$ (i.e. $\mathbf{I}=\mathrm{diag}(1,1,1,\ldots)$), $\mathbf{0}$ a matrix with all elements being 0, while $\mathds{1}$ a matrix with all elements being 1 (which should not be confused with $\mathbf{I}$).

\section{Problem Formulation} \label{sec:formulation}

 Given an $n$ qubit system, our goal is to find the optimal ${\color{black} 2^n } \times {\color{black} 2^n }$ transformation matrix $\mathbf{X}$ that transforms a given ${\color{black} 2^n } \times 1$ initial state vector $|\psi_\mathrm{initial} \rangle$ denoted as $\mathbf{A}$ into  the target ${\color{black} 2^n } \times 1$ final state vector $|\psi_{\mathrm{target}} \rangle$ denoted as $\mathbf{C}$. The optimality refers to the extend of sparsity: we intend that $\mathbf{X}$ is as sparse as possible so that quantum operations needed to realize it will be minimized.
 
  We are minimizing the distances between the obtained state $\mathbf{XA}$ and target $\mathbf{C}$, denoted by an objective function $F(\mathbf{X})$. This in turn solves $X |\psi_{\mathrm{initial}} \rangle = |\psi_{\mathrm{target}} \rangle$.
  Furthermore, one must impose constraints to $\mathbf{X}$. 
Firstly, $\mathbf{X}$ must be unitary, satisfying $\mathbf{X}^\dag \mathbf{X} = \mathbf{I}$, thus maintaining the probability conservation and reversibility of the quantum operations. 
Next, to ensure that the target states are properly normalized,  for each row or column vector of $\mathbf{X}$, denoted as $\mathbf{x_r}$ and $\mathbf{x_c}$ respectively, the sum of all its elements equals to one. Note that $r=1 \ldots {\color{black} 2^n }$ and $c=1 \ldots {\color{black} 2^n }$ represent $i$-th row and $j$-th column of $\mathbf{X}$. For ease of reference, we refer to these constraints as the `row/column-wise unit sum constraint' throughout the paper.

Thus, the objective function $F(\mathbf{X})$ for searching state transformation matrix $\mathbf{X}$ is given as:
\begin{equation} \label{eq:qst1}
F(X) = \frac{1}{2} \|\mathbf{X} \mathbf{A}-\mathbf{C}\|^2_2 \;\;, \;\;
\mathrm{s.t.} \;\; \mathbf{X}^\dag \mathbf{X} = \mathbf{I}  , \;\; \|\mathbf{x_{r}}\|_1 =1 ,  \;\; \|\mathbf{x_{c}}\|_1 = 1 \;\;,
\end{equation}
where ``s.t.'' is abbreviation for ``subject to'' and $\frac{1}{2}$ is added into the fidelity term to simplify subsequent optimization computations. 

The row/column-wise unit sum constraint can be equivalently expressed as an equality constraint in $F(\mathbf{X})$. 
In order to enforce row-wise unit sum, multiplication of objective $\mathbf{X}$ with a ${\color{black} 2^n } \times 1$ row $\mathds{1}_\mathrm{row}$ should equal  $\mathds{1}_\mathrm{row}$. Whereas column-wise unit sum can be represented as multiplication of column $1 \times {\color{black} 2^n }$ column vector $\mathds{1}_\mathrm{col}$ and $\mathbf{X}$ equal $\mathds{1}_\mathrm{col}$.


This reformulation enables us to represent the problem in the following form:

\begin{equation} \label{eq:qst2}
F(\mathbf{X}) = \frac{1}{2} \|\mathbf{X} \mathbf{A}-\mathbf{C}\|_2^2  \;\;,\;\;
\mathrm{s.t.} \;\; \mathbf{X}^\dag \mathbf{X} = \mathbf{I} \;\;, \mathbf{X} \mathds{1}_\mathrm{row} = \mathds{1}_\mathrm{row} \;\; ,\mathds{1}_\mathrm{col} \mathbf{X} = \mathds{1}_\mathrm{col} \;\;.
\end{equation}

Since objective matrix $\mathbf{X}$ is unitary, i.e. square matrix whose conjugate transpose is equal to its inverse, the rows and columns of $\mathbf{X}$ are orthonormal vectors. Eq.~\eqref{eq:qst2} can further be simplified, by merging two equality constraints into one. This property allows us to consider two equivalent constraints: $\mathds{1} \mathbf{X} = \mathds{1}$ and $\mathbf{X} \mathds{1} = \mathds{1}$, where $\mathds{1}$ represent ${\color{black} 2^n } \times {\color{black} 2^n }$ matrix of one. By multiplying $\mathbf{X}$ by $\mathds{1}$ on either side, we preserve the orthonormality of the rows and columns, ensuring that the resulting matrix satisfies the row/column-wise unit sum constraint.  Therefore, for the objective function $F(\mathbf{X})$, we employ $\mathds{1} \mathbf{X} = \mathds{1}$ to enforce the unit sum constraint for both rows and columns.

\begin{equation} \label{eq:qst3}
F(\mathbf{X}) = \frac{1}{2} \|\mathbf{X} \mathbf{A}-\mathbf{C}\|_2^2  \;\;,\;\;
\mathrm{s.t.} \;\; \mathbf{X}^\dag \mathbf{X} = \mathbf{I} \;\;, \mathds{1} \mathbf{X} = \mathds{1}.
\end{equation}

Next, we incorporate sparsity into the quantum state transformation problem represented by Eq.~\eqref{eq:qst3}. 
The sparsity constraint can be inserted into the optimization problem by introducing a penalty term with a Lagrange multiplier, denoted as $\lambda$. This Lagrange multiplier controls the balance between the sparsity of the transfermation matrix and the fidelity of the evolved state.

We now discuss the metric for sparsity. At first glance, $\|\cdot\|_{2,0}$ 
(i.e. the number of rows with nonzero elements) cannot be a good choice because the row/column-wise unit sum constraint precludes the possibility that elements in an entire row are zero.
To circumvent the problem, we impose either $\ell_{1}$ or $\ell_{2,1}$ sparsity given in Eq.~\eqref{eq:l1norm} and  Eq.~\eqref{eq:l21norm} into Eq.~\eqref{eq:qst3}. Both $\ell_{1}$ norm and $\ell_{2,1}$ norm are widely used as convex relaxations or approximations of the zero norm in the literature~\cite{anguluri2022localization, li2020hyperspectral, miao2019graph}, under which
the elements are encouraged to approach zero without necessarily becoming exactly zero.
To this end, we consider optimizing the following objective problems in the rest of the paper:
\begin{equation} \label{eq:l1QST}
F^{\ell_1}(\mathbf{X}) = \frac{1}{2} \|\mathbf{X} \mathbf{A}-\mathbf{C}\|_2^2 + \lambda \|\mathbf{X}\|_{1} \;\;, \;\;
\mathrm{s.t.} \;\; \mathbf{X}^\dag \mathbf{X} = \mathbf{I} \;\;, \mathds{1} \mathbf{X} = \mathds{1}.
\end{equation}

\begin{equation} \label{eq:l21QST}
\begin{split}
F^{\ell_{2,1}}(\mathbf{X}) = \frac{1}{2} \|\mathbf{X} \mathbf{A}-\mathbf{C}\|_2^2 + \lambda \|\mathbf{X}\|_{2,1} \;\;, \;\;
\mathrm{s.t.} \;\; \mathbf{X}^\dag \mathbf{X} = \mathbf{I} \;\;, \mathds{1} \mathbf{X} = \mathds{1}.
\end{split}
\end{equation}

Furthermore, our method extends beyond the specific objective function $F(\mathbf{X})$ that we have considered. It is capable of addressing sparse optimization problems with any convex objective function $\mathcal{F}(X)$ containing unitary constraint and equality constraint $\mathbf{D}\mathbf{X} = \mathbf{E}$.
Example of a widely employed convex objective function in compressed sensing is the least squares function $\mathcal{F}(\mathbf{X}) = \|\mathbf{A}\mathbf{X}-\mathbf{C}\|^2_2$,
Moreover, by carefully selecting an appropriate dictionary matrix $\mathbf{D}$  and target constraint matrix $\mathbf{E}$ \cite{gribonval2010dictionary}, we can represent various types of constraints.
For example, clustering constraint is represented by setting $\mathbf{D}$ as the matrix of cluster centroids and $\mathbf{E}$ as the matrix representing the assigned clusters for each data point~\cite{sprechmann2010dictionary}.
Generally, our algorithm can tackle problem with following form:
\begin{equation}\label{eq:general}
\min_\mathbf{X} \;\;\mathcal{F} (\mathbf{X}) + \lambda \|\mathbf{X}\|_{p} \;\; , \;\;
\mathrm{s.t.}  \;\; \mathbf{X}^\dag \mathbf{X}= \mathbf{I}  \;\; ,  \;\; \mathbf{D}\mathbf{X}= \mathbf{E} \;\;.
\end{equation}


\section{ADMM on Quantum State Transformation} \label{sec:method}
In this section we present our method for optimizing group-sparsed unitary matrix for $\mathbf{X}$ in Eq.~\eqref{eq:l21QST}.
The incorporation of unitary constraints in matrix optimization is particularly relevant in various quantum problems, including quantum channel estimation~\cite{chapeau2021quantum,chen2023unitarity}, quantum tomography~\cite{baldwin2014quantum, yang2017effective} and quantum state transformation~\cite{gleinig2021efficient, malvetti2021quantum, jaques2022leveraging, de2022double}.
{\color{black} Due to the nonconvexity of the unitary constraint, both of the optimization problem defined in Eq.~\eqref{eq:l1QST} and Eq.~\eqref{eq:l21QST} are NP-hard.}
 In this paper, we propose a new approach for solving the problem by introducing two dual variables, denoted as $\mathbf{Z_1}$ and $\mathbf{Z_2}$ and use ADMM algorithm. These dual variables assist us in effectively managing each constraint in the problem. Additionally, we introduce an auxiliary variable matrix $\mathbf{Y}$ that represents unitary constraint (i.e. $\mathbf{Y}^\dag \mathbf{Y} - \mathbf{I} = \mathbf{0}$). The final objective function to be optimized can be rewritten from Eq.~\eqref{eq:l21QST} to a new unconstrained objective function:
 
\begin{equation}  \label{eq:final}
\begin{split}
L(\mathbf{X}, \mathbf{Y}, \mathbf{Z_1}, \mathbf{Z_2}) = \frac{1}{2} \|\mathbf{X} \mathbf{A}-\mathbf{C}\|_2^2  + \lambda \|\mathbf{X}\|_{2,1} + \langle \mathbf{Z_1},\mathbf{X}-\mathbf{Y} \rangle +  \langle \mathbf{Z_2},\mathds{1}\mathbf{X}-\mathds{1} \rangle \\
+ \frac{\rho_1}{2} \|\mathbf{X}-\mathbf{Y}\|_2^2  + \frac{\rho_2}{2} \|\mathds{1} \mathbf{X}-\mathds{1}\|_2^2 \;\;,
\end{split}
\end{equation}
where $\langle \mathbf{\mathcal{A}},\mathbf{\mathcal{B}} \rangle$ represents the inner product of given matrices $\mathbf{\mathcal{A}}$ and $\mathbf{\mathcal{B}}$ and variables $\rho_1$ and $\rho_2$ are penalty parameters that control the balance between the quadratic penalty term and its corresponding dual variables $\mathbf{Z_1}$ and $\mathbf{Z_2}$.

Breaking down the unconstrained objective function $L$,
the first term represents the fidelity objective and the second term is the sparsity, the balance between which is controlled by $\lambda$. 
We use the row sparsity given by Eq.~\eqref{eq:l21norm} in the rest of this section, but our algorithm can be generally applied to other sparsity constraints.

Next, $ \langle \mathbf{Z_1},\mathbf{X}-\mathbf{Y} \rangle  +  \frac{\rho_1}{2} \|\mathbf{X}-\mathbf{Y}\|^2_F$, penalizes deviations from the unitary constraint and promotes the matrix to be closer to a unitary matrix.
Specifically, $\frac{\rho_1}{2} \|\mathbf{X}-\mathbf{Y}\|^2_F$ represents the Frobenius norm squared of the difference between $\mathbf{X}$ and $\mathbf{Y}$ weighted by $\rho_1$. 
Next, $ \langle \mathbf{Z_2},\mathds{1}\mathbf{X}-\mathds{1} \rangle + \frac{\rho_2}{2} \|\mathds{1} \mathbf{X}-\mathds{1}\|^2_F $ enforce each row and column of $\mathbf{X}$ sum up to 1.
Similarly, the second term represents the squared Frobenius norm of the difference between $\mathds{1} \mathbf{X}$ and $\mathds{1}$, weighted by $\rho_2$.
Since both the unitary constraint and the equality constraint fix the result $\mathbf{X}$ to meet the requirements of quantum state transformation, both constraints carry equal importance. Therefore, we take $\rho_1 = \rho_2 = \rho$.

By minimizing $L$ with respect to $\mathbf{X}$, $\mathbf{Y}$, $\mathbf{Z_1}$ and $\mathbf{Z_2}$, optimal transformation matrix  $\mathbf{X}$ that enhances the minimum values of the objective function while satisfying the sparsity and other structural requirements can be found.
{\color{black} The ADMM Algorithm for Quantum State Transformation provides a comprehensive outline for solving Eq.~\eqref{eq:final}.}

In order to ensure the convergence of our algorithm, we have implemented a termination condition based on the observation of the same value for 200 consecutive iterations. 
The algorithm will terminate once the condition is met to prevent unnecessary further iterations. 
This approach helps us maintain a balance between achieving convergence and avoiding excessive computation.

\begin{algorithm}
\caption{\bf ADMM Algorithm for Quantum State Transformation}
\label{admm_algorithm}
\begin{algorithmic}[1]
\REQUIRE Input parameters  $\lambda$ and $\rho$ and known matrix $\mathbf{A}$ and $\mathbf{C}$
\STATE Initialize variables: $\mathbf{X}^{(0)}$ to suitable initial values
\STATE Set $\mathbf{Y}^{(0)}$, $\mathbf{Z_1}^{(0)}$ and $\mathbf{Z_2}^{(0)}$ to zeros of the same size as $\mathbf{X}^{(0)}$
\STATE Normalize quantum state vectors $\mathbf{A}$ and $\mathbf{C}$.
\WHILE{not converged}
    \STATE Update $\mathbf{X}^{(k+1)}$ by minimizing the augmented Lagrangian with respect to $\mathbf{X}$:
    \begin{equation} \notag
	\begin{split}
	\mathbf{X}^{(k+1)} = \arg\min_\mathbf{X} \frac{1}{2} \|\mathbf{X}^k \mathbf{A}-\mathbf{C}\|_2^2  + \lambda \|\mathbf{X}^{k}\|_{2,1} + \langle \mathbf{Z_1}^{k} ,\mathbf{X}^{k} -\mathbf{Y}^{k} \rangle + \langle \mathbf{Z_2}^{k}, \mathds{1} \mathbf{X}^{k}-\mathds{1} \rangle \\
 + \frac{\rho_1}{2} (\|\mathbf{X}^{k} - \mathbf{Y}^{k}\|_2^2 + \frac{\rho_2}{2} \|\mathds{1} \mathbf{X}^{k} - \mathds{1}\|_2^2
       \end{split}
     \end{equation}
    \STATE Update unitary matrix $\mathbf{Y}^{(k+1)}$ by Singular Value Decomposition. 
    \STATE Update dual variables $\mathbf{Z_1}^{(k+1)}$ and $\mathbf{Z_2}^{(k+1)}$:
    \begin{equation} \notag 
		\mathbf{Z_1}^{k+1} = \mathbf{Z_1}^{k} + \rho_1 (\mathbf{X}^{k+1} - \mathbf{Y}^{k+1})
	\end{equation}
  \begin{equation}  \notag 
		\mathbf{Z_2}^{k+1} = \mathbf{Z_2}^{k} + \rho_2 (\mathds{1} \mathbf{X}^{k+1} - \mathds{1})
	\end{equation}
    \IF{convergence criteria met}
        \STATE Exit loop
    \ENDIF
\ENDWHILE
\end{algorithmic}
\end{algorithm}

\subsection{Updating X Sub-problem} 
In step 5 of {\color{black} our algorithm}, the focus is to update $\mathbf{X}$ by seeking its local optimum for each iteration through solving:
   \begin{equation} \label{eq:subprob}
	\begin{split}
	\mathbf{X}^{(k+1)} = \arg\min_\mathbf{X} \frac{1}{2} \|\mathbf{X}^k \mathbf{A}-\mathbf{C}\|_2^2  + \lambda \|\mathbf{X}^{k}\|_{2,1} + \langle \mathbf{Z_1}^{k} ,\mathbf{X}^{k} -\mathbf{Y}^{k} \rangle + \langle \mathbf{Z_2}^{k}, \mathds{1} \mathbf{X}^{k}-\mathds{1} \rangle \\
+ \frac{\rho_1}{2} (\|\mathbf{X}^{k} - \mathbf{Y}^{k}\|_2^2 + \frac{\rho_2}{2} \|\mathds{1} \mathbf{X}^{k} - \mathds{1}\|_2^2 \;\;,
       \end{split}
   \end{equation}
 by considering the proximal operator, which is an effective tool for addressing optimization problems involving non-smooth terms.

The proximal operator, denoted as $\text{prox}_{\lambda\|\cdot\|_{p}}(\mathbf{x})$, is defined as follows:
\begin{equation} \label{eq:proximal}
\text{prox}_{\lambda\|\cdot\|_{p}}(\mathbf{x}) = \arg\min_\mathbf{u} \left(\lambda \|\mathbf{u}\|_{p} + \frac{1}{2}\|\mathbf{u}-\mathbf{x}\|^2\right) \;\; .
\end{equation}

In Eq.~\ref{eq:proximal}, the variable $\mathbf{x}$ represents a given vector. The proximal operator is a mathematical operation that acts on $\mathbf{x}$ to produce a unique minimizer $\mathbf{u}$ by solving the optimization problem. The goal is to find the vector $\mathbf{u}$ that achieves the smallest value for the objective function in equation \eqref{eq:proximal} over all possible choices of $\mathbf{x}$. 


Specially, the proximal operator for $\ell_1$ penalty term is universally defined as~\cite{hu2017group,lou2018fast}:
\begin{equation}\label{softholding}
\text{prox}_{\lambda\|\cdot\|_{1}} \left(\mathbf{x}\right) = \text{sign}(\mathbf{x}_{ij}) \cdot \max \left(|\mathbf{x}_{ij}| - \lambda, 0\right) \;\;,
\end{equation}
and, the proximal operator for the $\ell_{2,1}$ norm is defined as follows:
\begin{equation}\label{prox_l21}
\text{prox}_{\lambda\|\cdot\|_{2,1}} \left(\mathbf{x}\right) =  \sum_{i=1}^m  \left( \sum_{j=1}^n  \text{sign}(\mathbf{x}_{ij}) \cdot \max(|\mathbf{x}_{ij}| - \lambda, 0)\right) \;\;.
\end{equation}

 The solution of Eq.~\eqref{eq:subprob} is given as $\mathbf{X} = \mathrm{prox}(\mathbf{\mathcal{X}}, \lambda)$, where $\mathbf{\mathcal{X}}^{k+1}$ is the local solution of Eq.~\eqref{eq:subprob}, equivalent to solving:
\begin{equation} \label{eq:xsub}
\begin{split}
\mathbf{\mathcal{X}}^{k+1} = \arg\min_\mathbf{X}  \frac{1}{2} \|\mathbf{X}^k \mathbf{A}-\mathbf{C}\|_2^2 + \langle \mathbf{Z_1}^{k} ,\mathbf{X}^{k} -\mathbf{Y}^{k} \rangle +  \langle \mathbf{Z_2}^{k},\mathds{1} \mathbf{X}^{k}-\mathds{1} \rangle \\
 + \frac{\rho_1}{2} \|\mathbf{X}^{k} - \mathbf{Y}^{k}\|_2^2 + \frac{\rho_2}{2} \|\mathds{1} \mathbf{X}^{k} - \mathds{1}\|_2^2 \;\;,
\end{split}
\end{equation}
with solution given $\mathbf{\mathcal{X}}^{k+1}$ as, 
\begin{equation}  \label{eq:subx_sol}
\mathbf{\mathcal{X}}^{k+1} = (\mathbf{A} \mathbf{A}^T + \rho \mathds{1}_N + \mathds{1}^T \mathds{1}) \ (\mathbf{C} \mathbf{A}' + \rho_1 \mathbf{Y} - \mathbf{Z_1} -\mathbf{Z_2} \mathds{1}^T - \mathds{1}^T \mathds{1}) \;\; .
\end{equation}

In general, for given convex $\mathcal{F}(\mathbf{X})$, we can solve for optimal $\mathbf{\mathcal{X}}^{k+1}$ with $\ell_{p}$ sparsity 
The solution of $\mathbf{X}$ is given as $X^{k+1} = \text{prox}(\mathbf{\mathcal{X}}^{k+1}, \lambda)$. 
where $\mathbf{\mathcal{X}}^{k+1}$ is the solution of $\arg\min_\mathbf{X} \mathcal{F}(\mathbf{X})$ with the constraints.

For example, in the case of an objective with least square form with unitary constraint, $\mathbf{\mathcal{X}}^{k+1}$ can be obtained by solving:
\begin{equation} \label{eq:xsub_ls}
\begin{split}
\mathbf{\mathcal{X}}^{k+1} = \arg\min_\mathbf{X}  \frac{1}{2} \|\mathbf{A} \mathbf{X}^{k}-\mathbf{C}\|_2^2 + \langle \mathbf{Z_1}^{k} ,\mathbf{X}^{k} -\mathbf{Y}^{k} \rangle +  \langle \mathbf{Z_2}^{k},\mathds{1} \mathbf{X}^{k}-\mathds{1} \rangle \\
 + \frac{\rho_1}{2} \|\mathbf{X}^{k} - \mathbf{Y}^{k}\|_2^2 + \frac{\rho_2}{2} \|\mathds{1} \mathbf{X}^{k} - \mathds{1}\|_2^2 \;\;,
\end{split}
\end{equation}
\begin{equation} \label{eq:xsub_ls2}
\mathbf{\mathcal{X}}^{k+1} = \arg\min_\mathbf{X} \frac{1}{2} \|\mathbf{A} \mathbf{X}^{k}-\mathbf{C}\|_2^2  + \langle \mathbf{Z_1}^{k} ,\mathbf{X}^{k} -\mathbf{Y}^{k} \rangle  + \frac{\rho_1}{2} \|\mathbf{X}^{k} - \mathbf{Y}^{k}\|_2^2 \;\;,
\end{equation}
either with or without row/column wise unit sum constraint. 


\subsection{Updating Y Subproblem}

Step 6 of {\color{black} the algorithm} aims to solve for $\mathbf{Y}$, which is the variable used to consider the unitary constraint. Since  $\mathbf{Y}$ is not involved in sparsity and row/column-wise unit sum constraint,  it can be treated as a sub-problem that search for an unitary matrices $\mathbf{Y}$ fulfilling the fidelity condition. Such problem can be considered as the Orthogonal Procrustes problem~\cite{schonemann1966generalized}.
The Orthogonal Procrustes problem aims to maximize $\langle \mathbf{X},\mathbf{C}\mathbf{A}^\dag \rangle$ which can be solved by Singular Value Decomposition, a matrix factorization routine that breaks down a matrix into three components: a left singular matrix $\mathbf{U}$, a diagonal singular value matrix $\mathbf{S}$, and a right singular matrix $\mathbf{V}$: 
\begin{equation} \label{SVD}
\mathbf{C}\mathbf{A}^\dag = \mathbf{U}\mathbf{S}\mathbf{V}^\dag
\end{equation} 

In the context of our Orthogonal Procrustes problem, the solution $\mathbf{Y}$ can be obtained by setting $\mathbf{Y} = \mathbf{U}\mathbf{V}^\dag$. This choice of $\mathbf{Y}$ maximizes the inner product and represents the optimal solution to the problem.

\section{Construction of Quantum Circuit using Optimal Unitary Matrix}\label{sec:construction}

To ensure the effectiveness of the circuit implementation, it is crucial to obtain the optimal unitary matrix $\mathbf{X}$ using ADMM from the above section before proceeding with the construction of the quantum circuit. 
The ADMM algorithm determined a pre-specified transformation matrix $\mathbf{X}$ that facilitates the transformation from the initial state $\mathbf{A}$ to the final state $\mathbf{C}$.
This ensures that the transformation achieved by the circuit aligns as closely as possible with the desired outcome. 
In this section, we will discuss the process of constructing a quantum circuit using the previously obtained optimal unitary matrix $\mathbf{X}$.

Based on the Solovay-Kitaev theorem, which stated that any unitary transformation can be approximated with arbitrary precision using a sequence of single-qubit gates and the Controlled-NOT (CNOT) gate.
Various decomposition methods have been proposed to efficiently break down unitary transformation matrices into sequences of elementary gates such as methods proposed in 
\cite{malvetti2021quantum, mottonen2004quantum, li2013decomposition, li2014decomposition, di2013synthesis, geller2021experimental}. 
In this paper, we propose to convert the obtained optimal transformation matrix $\mathbf{X}$ into a circuit representation by following the steps outlined in~\cite{li2013decomposition, li2014decomposition}.

\begin{enumerate}

\item \textbf{Two-level decomposition:} 
Firstly, we employ the cosine-sine decomposition method~\cite{mottonen2004quantum, di2013synthesis} to decompose the obtained $2^n \times 2^n$ transformation matrix $\mathbf{X}$ into a product of a series of $2^n \times 2^n$ block-diagonal matrices $\mathbf{U}_k$, $k = 1, 2, 3... 2^n$.
This technique involves breaking down the unitary matrix into a product of simpler block-diagonal matrices, where each block corresponds to a smaller unitary matrix $\mathbf{u}_k$.
\begin{equation} \label{decompose}
\mathbf{X} = \mathbf{U}_1 \mathbf{U}_2 \mathbf{U}_3 \ldots \mathbf{U}_n \;\; ,
\end{equation}

\item \textbf{Circuit Formation:}
To represent transformation matrix $\mathbf{X}$ as a circuit, we associate each decomposed block-diagonal matrix $\mathbf{U}_k$ with a specific gate $G_k$.
We break the process of circuit formation into two steps:
\begin{enumerate}
    \item \textbf{Qubit identification:}
    First, the position of each $2 \times 2$ unitary matrix $\mathbf{u}_k$ within $\mathbf{U}_k$ indicates the qubits on which the corresponding gate operates. 
Through examining the positions of each $\mathbf{u}_k$, we can identify the corresponding qubits transformation matrix $\mathbf{U}_k$ act on.

In a quantum system, the transformation matrix represents the evolution of quantum states. Each entry within the matrix corresponds to the transition amplitude between specific quantum states.
This information is crucial for correctly mapping the gates onto the qubits in the quantum circuit.
When $\mathbf{u}_k$ is located at positions $(i,i)$, $(i, j)$, $(j, i)$ and $(j,j)$, the fact indicates that the amplitudes for transitioning between the  $i$-th and $j$-th input states corresponds to the $i$-th and $j$-th  output states, respectively. 
Ref.~\cite{li2014decomposition} provides a list of possible gate choices for two to four qubits. 
For example, in a $3$-qubit system, the element at position $(i, j)$ in the unitary matrix corresponds to the transition amplitude from state $|j-1\rangle$ to state $|i-1\rangle$ depending on the position of the $\mathbf{u}_k$, where the transformation can be represented by an $8 \times 8$ matrix:

\begin{equation}  \label{eq:3x3matrix}
\begin{aligned}
\begin{array}{c|cccccccc}
    & |000\rangle & |001\rangle & |010\rangle & |011\rangle & |100\rangle & |101\rangle & |110\rangle & |111\rangle \\
\hline
|000\rangle & x_{1,1} & x_{1,2} & x_{1,3} & x_{1,4} & x_{1,5} & x_{1,6} & x_{1,7} & x_{1,8} \\
|001\rangle & x_{2,1} & x_{2,2} & x_{2,3} & x_{2,4} & x_{2,5} & x_{2,6} & x_{2,7} & x_{2,8} \\
|010\rangle & x_{3,1} & x_{3,2} & x_{3,3} & x_{3,4} & x_{3,5} & x_{3,6} & x_{3,7} & x_{3,8} \\
|011\rangle & x_{4,1} & x_{4,2} & x_{4,3} & x_{4,4} & x_{4,5} & x_{4,6} & x_{4,7} & x_{4,8} \\
|100\rangle & x_{5,1} & x_{5,2} & x_{5,3} & x_{5,4} & x_{5,5} & x_{5,6} & x_{5,7} & x_{5,8} \\
|101\rangle & x_{6,1} & x_{6,2} & x_{6,3} & x_{6,4} & x_{6,5} & x_{6,6} & x_{6,7} & x_{6,8} \\
|110\rangle & x_{7,1} & x_{7,2} & x_{7,3} & x_{7,4} & x_{7,5} & x_{7,6} & x_{7,7} & x_{7,8} \\
|111\rangle & x_{8,1} & x_{8,2} & x_{8,3} & x_{8,4} & x_{8,5} & x_{8,6} & x_{8,7} & x_{8,8} \\
\end{array}
\end{aligned}
\end{equation}

    \item \textbf{Gate Construction:}
To construct each gate $G_i$, $\mathbf{u}_k$ can be extracted from the corresponding $\mathbf{u}_k$ and represented using a sequence of rotational matrices, which can be implemented as single-qubit gates. 
By representing each $\mathbf{U}_k$ gate within the circuit as a series of rotations, the matrix for transformation from state $\mathbf{A}$ into state $\mathbf{C}$ in the quantum system can be represented purely by a sequence of rotations. These rotations encode the transitions between specific quantum states, enabling the construction of the desired circuit representation for $\mathbf{X}$. 

Specifically, each $\mathbf{U}_k$ can be decomposed into a product of rotational matrices $R_z(\theta)$, $R_y(\phi)$, and $R_z(\lambda)$, with Euler angles $\theta$, $\phi$ and $\lambda$, by considering each $\mathbf{u}_k$ gates as:

\begin{equation} \label{eq:rotation}
\mathbf{u}_k = \begin{bmatrix}
\alpha & \beta \\
\gamma & \delta 
\end{bmatrix}
= R_z(\theta) \cdot R_y(\phi) \cdot R_z(\lambda) \;\; ,
\end{equation}
implying that  $\mathbf{U}_k$ contains elements $\left( i,i \right) = \alpha$, $\left( i,j \right) = \beta$, $\left( j,i \right) = \gamma$ and $\left( j,j \right) = \delta$.
This decomposition allows us to understand the individual rotational components that contribute to the overall transformation represented by the matrix $\mathbf{U}_k$.
By applying these rotational matrices in the specified order, the desired gate $G_i$ is constructed.

\end{enumerate}
These two steps enable the translation of block diagonal matrices $\mathbf{U}_1, \mathbf{U}_2, \mathbf{U}_3, \ldots, \mathbf{U}_k$ into their corresponding gates $G_1, G_2, G_3, \ldots, G_k$ within the quantum circuit representation.

\item \textbf{Optimization:} 
One may further optimize the circuit by merging or cancelling adjacent gates on the same pair of qubits, which help reducing the total number of gates and improve the overall performance of the circuit.

\end{enumerate}
By following these steps, we can construct a quantum circuit using the optimal transformation matrix $\mathbf{X}$ obtained through optimization. 
This circuit enables the transformation of quantum states according to the desired transformation represented by $\mathbf{X}$.

\section{Numerical simulation} \label{sec:results}
In this section, we present numerical simulations conducted on our proposed ADMM method for sparse and group sparse Quantum State Transformation. The purpose of these simulations is to demonstrate the performance of our algorithm.

We generate a collection of complex vectors with size ${\color{black} 2^n } \times 1$ to simulate an entangled state of $n$ qubits. This simulation replicates the process of searching for a transformation matrix that represents a quantum circuit capable of transforming an initial state $\mathbf{A}$ to a target state $\mathbf{C}$, both randomly generated in MATLAB. Additionally, we set the initial transformation matrix $\mathbf{X}_0$ as a zero matrix $\mathbf{0}$, recognizing that the initial variables can significantly impact ADMM optimization.

The simulation section is divided into two parts. The first part aims to illustrate the performance of the proposed algorithm under various parameters and scenarios, analyzing its behavior and effectiveness in different settings. Each data point in the tables represents an average of 30 trials.
The second part provides an example of a $3$-qubit system, demonstrating how to generate the transformation and the corresponding quantum circuit given the initial and final states.
All numerical simulations are implemented in MATLAB R2021a and executed on a personal computer with an Intel(R) Core(TM) i7-8750H CPU clocked at 2.21 GHz and 16 GB RAM.


\subsection{Convergence of the Proposed Algorithm}

\begin{table}
\centering
    \begin{tabular}{|c|c|c|c|}
    \hline
    \multicolumn{4}{|c|}{${\ell_{1}}$-Norm}\\
    \hline
    \textbf{Number of Qubits} &  n = 2 & n = 3 &  n = 4  \\
    \hline
    {Converged value of $L$} & $1.155$ & $1.195$ & $1.439$ \\
    \hline
    {No. of iterations required for convergence} & $>1000$ & $>1000$ & $>1000$ \\
    \hline
    {Percentage of Non-Zero Elements} & $<78 \%$ & $<74 \%$ & $<60 \%$ \\
    \hline
    \end{tabular}
    
    \vspace{10pt}

    \begin{tabular}{|c|c|c|c|c|c|c|}
    \hline
    \multicolumn{7}{|c|}{${\ell_{2,1}}$-Norm}\\
    \hline
    \textbf{Number of Qubits} &  n = 2 & n = 3 &  n = 4 &  n = 5 &  n = 6 &  n = 7 \\
    \hline
    {Converged value of $L$} & $1.269$ & $1.475$ & $1.923$ & $2.633$ & $3.663$ & $4.861$ \\
    \hline
    {No. of iterations required for convergence} & $>100$ & $>100$ & $>100$ & $>100$ & $>100$ & $>100$ \\
    \hline
    {Number of Non-Zero Elements} & $16$ & $64$ & $256$ & $1024$ & $4096$ & $16384$  \\
    \hline
    \end{tabular}
\caption{Comparison of ${\ell_1}$ and $\ell_{2,1}$ norm on randomly generated vectors $\mathbf{A}$ and $\mathbf{C}$.}
\label{tab:QST_compareL}
\end{table}

This subsection presents simulations carried out on our proposed ADMM method for sparse quantum state transformation. A sufficient number of simulations were performed to ensure the convergence of our algorithm in searching for transformation matrices with introduced sparsity. Table~\ref{tab:QST_compareL} compares the $\ell_1$ and $\ell_{2,1}$ sparsity norms for the same datasets, considering different numbers of qubits. Additionally, Appendix~\ref{appx:appxA} provides an analysis of the sparsity norms without vector normalization. Furthermore, similar simulations were conducted on least square objective problems, and detailed results can be found in Appendix~\ref{appx:appxB}. These results demonstrate the efficient convergence of our algorithm within a limited number of iterations and time.

\subsubsection{Sparsity Comparison}
In these simulations, we investigated the impact of sparsity on the quantum state transformation problem, specifically focusing on the effects of the $\ell_{1}$ norm and $\ell_{2,1}$ norm penalties. Our findings consistently indicated that, for normalized vectors, the $\ell_{1}$ norm outperformed the $\ell_{2,1}$ norm in terms of achieving superior optimal values and generating a more sparse transformation matrix, as demonstrated in Table~\ref{tab:QST_compareL} and Figure~\ref{fig:X_val}. The consistent superiority of the $\ell_{1}$ norm penalty for normalized vectors, in achieving better optimal solutions while satisfying the given constraints, highlights the effectiveness of this approach in promoting sparsity and facilitating the search for sparse transformation matrices in quantum state transformation. 

However, it is important to note that as the number of qubits increases, the $\ell_1$ norm becomes less suitable as it tends to converge quickly to an identity matrix, even with small penalty values of $\rho$. This limitation arises due to the normalization requirement that the sum of the squares of all elements in the vector be equal to $1$. As the number of elements in the vector increases, each element in the normalized vector tends to have a smaller value. Consequently, the penalty values become larger relative to the magnitude of each element in the vector. These larger penalty values impose overly restrictive constraints on unitarity and row/column-wise unit sums, leading to the algorithm's failure to achieve an acceptable fidelity. As a consequence, the resulting matrix becomes dominated by the identity structure, diminishing the effectiveness of the transformation. 

In such cases, it is advisable to consider the $\ell_{2,1}$ norm as an alternative regularization approach.The $\ell_{2,1}$ norm encourages row sparsity in the matrix $\mathbf{X}$, resulting in a more evenly distributed set of values across each row, as depicted in Figure~\ref{fig:X_val}b. On the contrary, the $\ell_{1}$ norm tends to produce a combination of higher-valued elements and sparser elements within the rows. This distinction arises from the mathematical properties of these sparsity terms, with the $\ell_{2,1}$ norm promoting a greater number of rows with near-zero values, while the $\ell_{1}$ norm encourages a mix of zero and non-zero values within each row.

Furthermore, in Appendix~\ref{appx:appxA}, we extended our investigation to consider unnormalized vectors in the same experiment. The results, along with Appendix~\ref{appx:appxB} on least square optimization problems, consistently showed that the $\ell_{2,1}$ norm penalty outperformed the $\ell_{1}$ norm in terms of achieving better optimal solutions while satisfying the given constraints. This observation is attributed to the characteristics of unnormalized vectors having different magnitudes of elements. Due to these variations, the $\ell_{2,1}$ norm penalty demonstrates better performance compared to the $\ell_{1}$ norm penalty in handling the differences in magnitudes and achieving better optimal solutions. However, when we normalize the vectors, bringing them to a similar scale, the $\ell_{1}$ norm penalty becomes more sensitive to individual element magnitudes. 
Thus, the observed data provides a compelling explanation for the superior performance of the $\ell_{2,1}$ norm penalty for unnormalized vectors and better performance for $\ell_{1}$ norm penalty after normalization.

\begin{figure} 
  \centering
(a)\includegraphics[width=0.68\columnwidth]{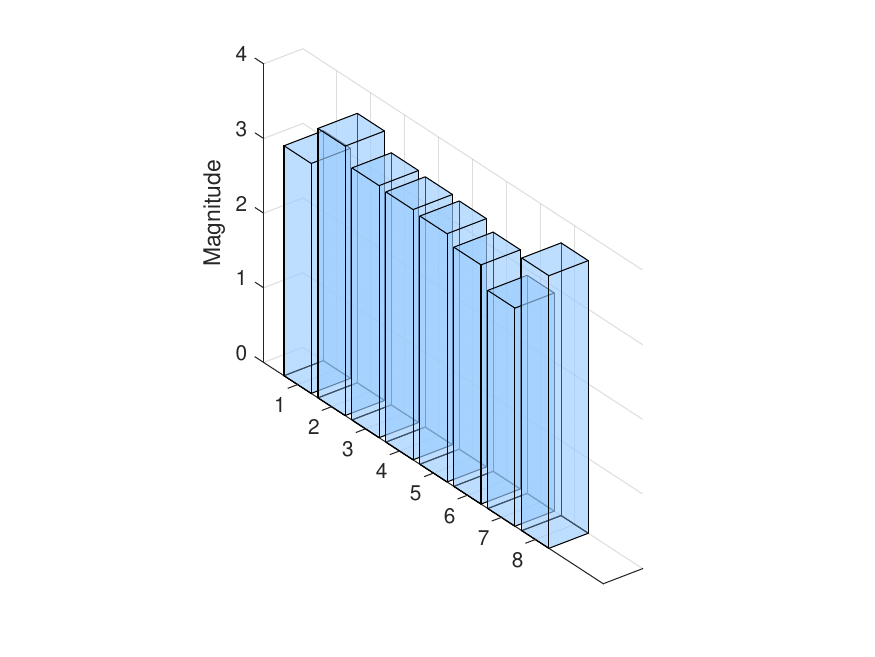}
(b)\includegraphics[width=0.68\columnwidth]{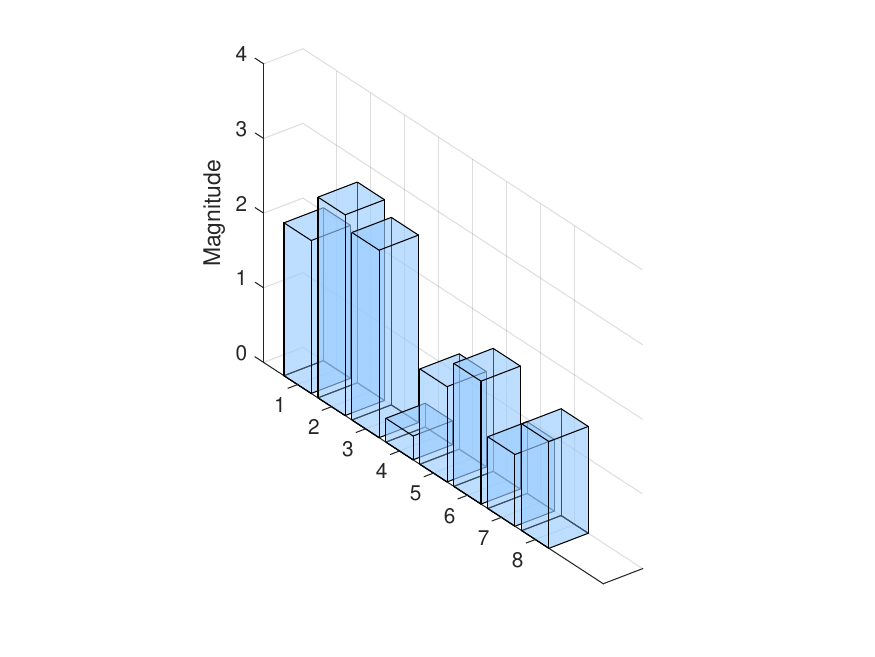}
  \caption{Magnitude of sparsed values in each row of $\mathbf{X}$ by (a) $\ell_1$-norm and  (b) ${\ell_{2,1}}$-norm.}
  \label{fig:diff_row}
\end{figure}

Importantly, when the goal is to obtain a more sparse approximation of a small-sized transformation matrix or address other optimization problems with similar constraints (as shown in Appendix~\ref{appx:appxA} and Appendix~\ref{appx:appxB}), the $\ell_1$ norm is more suitable than the $\ell_{2,1}$ norm, as indicated in Figure~\ref{fig:diff_row}. Comparing our optimal results to the transformation matrix solution obtained without introducing sparsity, we observe that the $\ell_1$ norm yields a higher magnitude of sparsity within each row compared to the $\ell_{2,1}$ norm.

These observations highlight the trade-offs and considerations involved in selecting the appropriate sparsity norm for optimization problems. Each norm offers advantages in specific contexts where sparsity is a desirable property. By effectively promoting the desired sparsity patterns, the sparsity term enhances the performance of accurate quantum state transformation and least square objectives, while adhering to unitary and row/column-wise unit sum constraints.


\subsubsection{Limitations}
Based on the simulation results, we conducted a comprehensive analysis of the convergence performance of our algorithm. Notably, as the number of qubits increased to $6$, as indicated in Table~\ref{tab:QST_compareL}, there was a discernible exponential growth in the required number of iterations to reach the optimal point.

Despite the observed exponential increase in the number of iterations, we consistently found that for cases involving $7$ or fewer qubits, the average time taken for each iteration remained relatively low, consistently below $0.03$ seconds. 
However, when the number of qubits increased to $8$, corresponding to state vectors of size $64 \times 1$ in dimension, the average time required for each iteration significantly increased to $0.025$ seconds, and achieving convergence demanded over $5000$ iterations. Furthermore, for systems with $9$ qubits, the average iteration time further increased to $0.095$ seconds.
These findings indicate that our algorithm is limited in its capability to search for unitary matrices for quantum state transformations up to $7$ qubits, regardless of the chosen sparsity constraint.
 
 Another limitation of our algorithm is the approximate setting of each row and column to one, rather than achieving an exact value of one. While we impose the constraint of summing each row and column to approximately one, there may exist a small deviation from the desired exact value. This approximation may introduce a slight deviation in the preservation of state probability, potentially affecting the accuracy of the quantum state transformation.

Although it is possible to impose the constraint such that each row and column exactly sums to one by setting a very high value for $\rho_2$, this approach may diminish the importance of other constraints and introduce trade-offs in the optimization process. Therefore, finding alternative methods to enforce the constraint of exact row and column sums equal to one in a more precise manner would be valuable in future research. This would ensure a more accurate preservation of state probability and enhance the reliability of the transformation process, leading to more precise and accurate outcomes.



\subsection{Example: Quantum circuit for transformation of a 3-qubit system}
In this subsection, we present a practical example of a $3$-qubit system to demonstrate the effectiveness of our algorithm. 
We showcase how our approach can generate the necessary transformation matrix $\mathbf{X}$ and design a corresponding quantum circuit to achieve the desired state transition. 
This example serves as a concrete illustration of the algorithm's capability in real-world quantum state transformation problems.

\subsubsection{Calculation of Transformation Matrix}

Firstly, we randomly generated the initial state $\mathbf{A} =|\psi_{\mathrm{initial}}\rangle$ and target state $\mathbf{C}=|\psi_{\mathrm{target}}\rangle$ of three qubits system:

\begin{equation} \label{AC_data}
\begin{aligned}
\mathbf{A} &= \begin{bmatrix}
 -0.2123 - 0.2389i \\
  -0.1660 + 0.2091i \\
  -0.1137 + 0.0785i \\
  -0.0718 - 0.0065i \\
   0.0028 + 0.0409i \\
  -0.6785 - 0.3505i \\
  -0.1024 - 0.0189i \\
   0.2783 + 0.3593i
\end{bmatrix} &
\mathbf{C} = \begin{bmatrix}
   0.0223 + 0.4588i \\
   0.0094 - 0.5120i \\
  -0.1664 + 0.5054i \\
  -0.0070 + 0.0765i \\
   0.0527 + 0.2267i \\
   0.0967 - 0.3773i \\
  -0.0845 - 0.1338i \\
  -0.0536 - 0.0630i
\end{bmatrix}
\end{aligned}
\end{equation}

\begin{figure} 
  \centering
(a)\includegraphics[width=0.68\columnwidth]{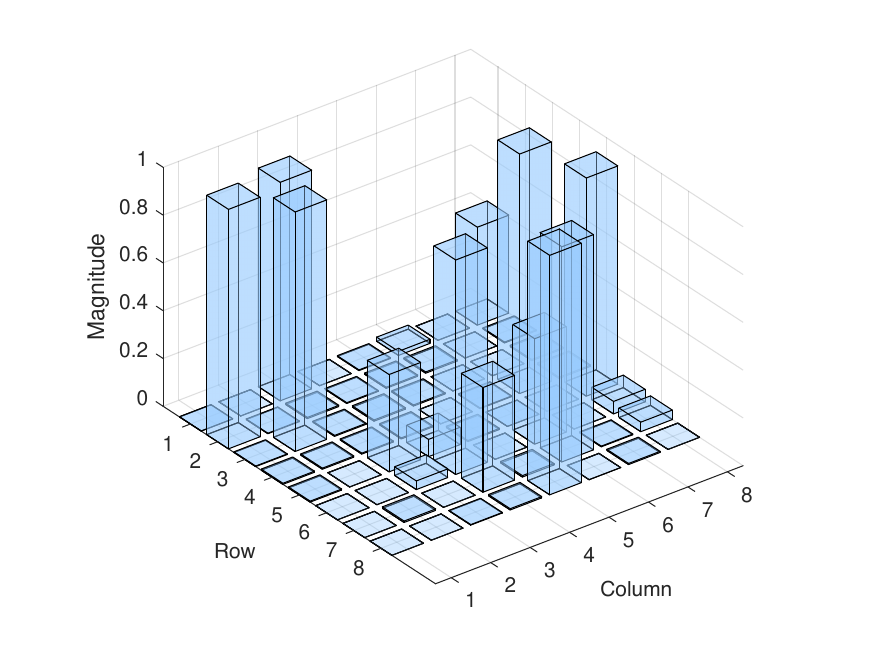}
(b)\includegraphics[width=0.68\columnwidth]{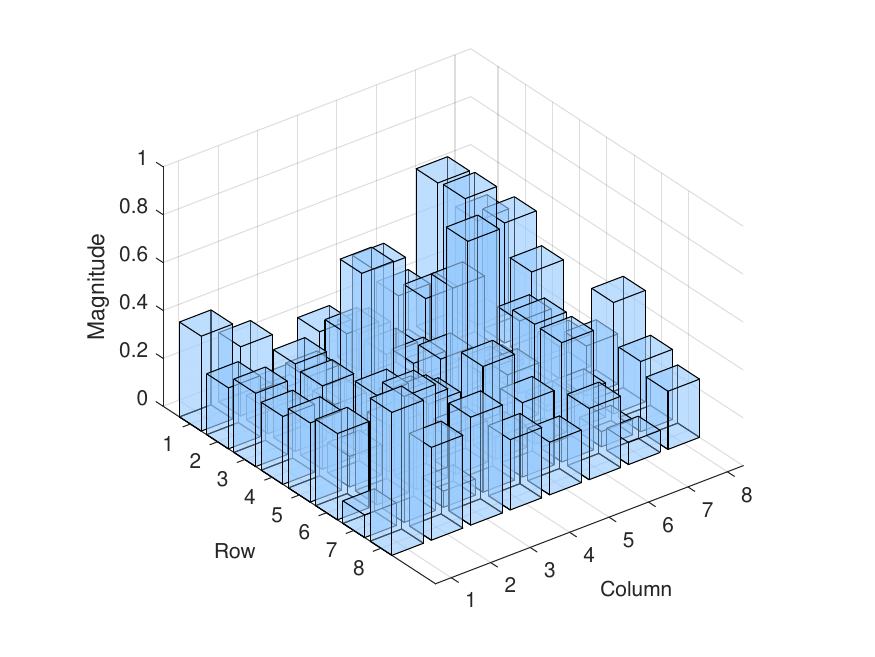}
  \caption{Magnitude of the resulted matrix (a) $\mathbf{X}_{\ell_1}$ and  (b) $\mathbf{X}_{\ell_{2,1}}$.}
  \label{fig:X_val}
\end{figure}

By employing the ADMM algorithm to minimize our objective function~\eqref{eq:final} with normalized $\textbf{A}$ and normalized $\textbf{C}$ given in Eq.~\eqref{AC_data}, we obtain the specific transformation matrix $\mathbf{X}$ for these states, denoted as $\mathbf{X}_{\ell_1}$ and $\mathbf{X}_{\ell_{2,1}}$. Each representing the transformation matrix to transform state $\mathbf{A}$ into state $\mathbf{C}$, with the introduction of sparsity constraint of $\ell_1$ norm for $\mathbf{X}_{\ell_1}$ and $\ell_{2,1}$ norm for $\mathbf{X}_{\ell_{2,1}}$. The magnitude of each element within the obtained matrices are shown as 3D plots in Fig.~\ref{fig:X_val}.
The resulting converged values of $L$ yields 1.0343 for $\mathbf{X}_{\ell_1}$ and 1.3849 for $\mathbf{X}_{\ell_{2,1}}$ norm.
These outcomes align with our findings from comparing the $\ell_1$ and $\ell_{2,1}$ norms, demonstrating that $\ell_1$ sparsity can yield superior optimal points. 
Moreover, both $\mathbf{X}_{\ell_1}$ and $\mathbf{X}_{\ell_{2,1}}$ are unitary matrices that approximately satisfy the property of having the sum of each row and column equal to 1. This characteristic ensures the preservation of total probability within the quantum circuit.

\subsubsection{Construction of Quantum Circuit}
In the remaining part of this section, we aim to construct the corresponding circuits for $\mathbf{X}_{\ell_1}$ and $\mathbf{X}_{\ell_{2,1}}$, following the procedures outlined in Section~\ref{sec:construction}.

\begin{figure} 
  \centering
\includegraphics[width=0.9\columnwidth]{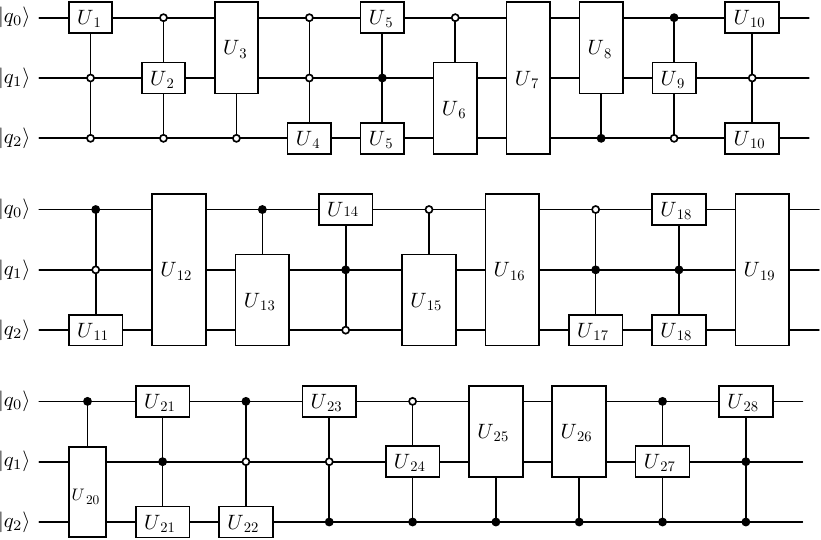}
  \caption{Quantum circuit for transformation of a 3-qubit system.}
  \label{fig:circuit}
\end{figure}

For a three-qubit system, the transformation matrix can be decomposed into 28 matrices, denoted as $U_{1}$ to $U_{28}$. The implementation of the three-qubit unitary matrix is visually represented by the quantum circuit diagram in Figure~\ref{fig:circuit}. This circuit diagram displays the arrangement and interconnections of the gates $U_1$ to $U_{28}$, offering a visual depiction of the quantum computation process. Each of these gates plays a vital role in capturing specific aspects of the transformation process and is essential for the successful implementation of the desired quantum operations.

To provide detailed insights into each of the $\mathbf{U}_k$ gate, we provide comprehensive information about their corresponding decomposed matrix $\mathbf{u}_k$ in Tables~\ref{tab:info_UL1} and~\ref{tab:info_UL21} for $\mathbf{X}_{\ell_1}$ and $\mathbf{X}_{\ell_{2,1}}$, respectively. These tables include the positions of the gate elements within $\mathbf{U}_k$ and the values of those elements ($\alpha$, $\beta$, $\gamma$, and $\delta$). 

Furthermore, each $\mathbf{u}_k$ gate can be further decomposed into a series of rotational matrices denoted as $R(\theta)$, $R(\phi)$, and $R(\lambda)$, as shown in Eq.~\eqref{eq:rotation}. These rotational gates play a crucial role in achieving the desired transformations. To provide precise details, we provide the specific angles for the rotational gates representing each $\mathbf{U}_k$ gate in Tables~\ref{tab:angle_UL1} and~\ref{tab:angle_UL21} for the optimal $\ell_1$ and $\ell_{2,1}$ transformation matrices, respectively. These angles serve as the instructions for the necessary rotations required for the respective transformations.

In summary, the construction of the quantum circuit for transforming the initial state to the target state of a three-qubit system has been successfully accomplished. Building upon the transformation matrix obtained using the ADMM method from the previous section, we proceeded with decomposing the matrix into individual gates. By carefully analyzing the corresponding matrices and elements, we determined the precise angles for the rotational gates. 
Through this construction, we have provided a way to simplify the execution of quantum operations on the three-qubit system based on sparse matrices.

\subsubsection{Gate Count for Transformation Matrix Reconstruction}

The sparsity of the transformation matrix plays a crucial role in determining the number of gates required for the resulting decomposed matrix. In our simulations, we observed that certain $U$ gates can be represented by the identity matrix and omitted, resulting in a reduction in the total number of gates. Specifically, for the example of decomposing the $\mathbf{X}_{\ell_1}$ transformation matrix, we found that $U_{5}$, $U_{6}$, $U_{7}$, $U_{11}$ and $U_{13}$ can be discarded, leading to a reduction of two gates in the decomposition process for the $\mathbf{X}_{\ell_1}$ transformation matrix.

In addition, we explored gate decomposition for cases with larger number of qubits with $\ell_1$ norm. These simulations specifically involved decomposing the transformation matrix $\mathbf{X}$ obtained using our ADMM algorithm. For a more comprehensive understanding of these simulations, please refer to Appendix~\ref{appx:appxD}.
 
By introducing sparsity into the transformation matrix before the decomposition, we create a matrix structure where zero elements correspond to negligible transitions between states. This characteristic allows us to achieve a significant reduction in the number of gates required for the decomposition process. Consequently, leveraging sparsity enables us to optimize gate utilization and minimize the resource overhead associated with quantum circuits.

\begin{table}[p]
\centering
\begin{tabular}{ |c|c|c|c|} 
\hline
$U$ & Position in $\mathbf{u}_k$  & $\alpha$, $\delta$ & $\beta$, $\gamma$ \\
\hline
$U_{1}$ & $(1,1)$ $(1,2)$ $(2,1)$ $(2,2)$  & $\pm 0.0007 - 0.0005i$  &$ 0.9944 \mp 0.1052i$ \\
\hline
$U_{2}$ & $(1,1)$ $(1,3)$ $(3,1)$ $(3,3)$  & $\pm 1$ & $0.0024 \pm 0.0006i$ \\
\hline
$U_{3}$ & $(1,1)$ $(1,4)$ $(4,1)$ $(4,4)$ & $\pm 1$ &  $0.0044 \mp 0.0004i$ \\
\hline
$U_{4}$ & $(1,1)$ $(1,5)$ $(5,1)$ $(5,5)$   & $\pm 1$ &  $0.0022 \pm 0.0034i$\\
\hline
$U_{5}$ &  $-$  & $-$ & $-$ \\
\hline
$U_{6}$ &  $-$  & $-$ & $-$   \\
\hline
$U_{7}$ & $-$  & $-$ & $-$ \\
\hline
$U_{8}$ & $(2,2)$ $(2,3)$ $(3,2)$ $(3,3)$ &  $\mp 0$ &  $ -0.9838 \mp  0.1793i$ \\
\hline
$U_{9}$ &  $(2,2)$ $(2,4)$ $(4,2)$ $(4,4)$ &  $\pm 1$ & $-0.0044 \mp 0.0008$ \\
\hline
$U_{10}$ &  $(2,2)$ $(2,5)$ $(5,2)$ $(5,5)$  &  $\pm 1$ & $0 \mp 0i$ \\
\hline
$U_{11}$ & $-$  & $-$ & $-$   \\
\hline
$U_{12}$ &  $(2,2)$ $(2,7)$ $(7,2)$ $(7,7)$  &  $\pm 1$ & $0.0051 \pm 0.0028 $ \\
\hline
$U_{13}$ &  $-$  & $-$ & $-$ \\
\hline
$U_{14}$ &  $(3,3)$ $(3,4)$ $(4,3)$ $(4,4)$  & $\mp  -0.7605 - 0.6493i$ & $ 0.0042  \pm 0.0007i$ \\
\hline
$U_{15}$ &  $(3,3)$ $(3,5)$ $(5,3)$ $(5,5)$  & $\pm 0.9135$ &  $0.0794 \mp 0.3990i$ \\
\hline
$U_{16}$ &  $(3,3)$ $(3,6)$ $(6,3)$ $(6,6)$  & $\pm 0.9994$  & $0.0237 \pm 0.0272i$i \\
\hline
$U_{17}$ &  $(3,3)$ $(3,7)$ $(7,3)$ $(7,7)$  & $\pm 1$ & $0 \mp 0i$\\
\hline
$U_{18}$ &  $(3,3)$ $(3,8)$ $(8,3)$ $(8,8)$  & $\pm 1$ & $ 0.0020 \pm 0.0006i$\\
\hline
$U_{19}$ &  $(4,4)$ $(4,5)$ $(5,4)$ $(5,5)$  &  $\pm 0.0451 + 0.0495i$ & $ -0.2892 \mp 0.9549i$ \\
\hline
$U_{20}$ &  $(4,4)$ $(4,6)$ $(6,4)$ $(6,6)$  &  $\pm 0.0737 + 0.0000i$ & $-0.9233 \mp 0.3771i$ \\                               
\hline
$U_{21}$ &  $(4,4)$ $(4,7)$ $(7,4)$ $(7,7)$  &   $\pm 0.8992 + 0.0000i$ & $ -0.0798 \mp 0.4301i$ \\ 
\hline
$U_{22}$ &  $(4,4)$ $(4,8)$ $(8,4)$ $(8,8)$  &  $\pm 1$ &  $ -0.0050 \mp 0.0015i$ \\
\hline
$U_{23}$ &  $(5,5)$ $(5,6)$ $(6,5)$ $(6,6)$  &  $\pm 0.0135 + 0.0738i $& $-0.9104 \pm 0.4069i$ \\
\hline
$U_{24}$ &  $(5,5)$ $(5,7)$ $(7,5)$ $(7,7)$  &   $\pm 0.6668$ & $-0.6491 \mp 0.3660i$ \\
\hline
$U_{25}$ &  $(5,5)$ $(5,8)$ $(8,5)$ $(8,8)$  & $\pm  0.0074 $ &  $0.9903 \pm 0.1386i$ \\
\hline
$U_{26}$ &  $(6,6)$ $(6,7)$ $(7,6)$ $(7,7)$  &  $\mp 0.0030 + 0.0024i $ & $  0.9966 \pm 0.0829i$\\
\hline
$U_{27}$ &  $(6,6)$ $(6,8)$ $(8,6)$ $(8,8)$  &  $\pm 0 0.6953$ & $  0.6901 \mp 0.2007i$\\
\hline
$U_{28}$ &  $(7,7)$ $(7,8)$ $(8,7)$ $(8,8)$  & $\alpha = -0.0832 + 0.9965i  $ & $\beta =  0.0106 - 0.0028i$ \\
  &    &  $\gamma = -0.0020 + 0.0108i$ & $\delta = -0.9904 + 0.1570i$ \\
\hline
\end{tabular}
\caption{Details of sub-matrices $\mathbf{u}_k$ for $X_{\ell_1}$}
\label{tab:info_UL1}
\end{table}
\clearpage
\begin{table} [p]
\centering
\begin{tabular}{|c|c|c|c|} 
\hline
U & Position in Matrix &  $\alpha$, $\delta$ & $\beta$, $\gamma$ \\
\hline
$U_{1}$ & $(1,1)$ $(1,2)$ $(1,8)$ $(2,2)$ & $\mp 0.8216 + 0.1820i $&$  0.5351 \mp 0.0748i$ \\
\hline
$U_{2}$ & $(1,1)$ $(1,3)$ $(3,1)$ $(3,3)$ & $\pm 0.8384$  & $ 0.5415 \pm 0.0631i$  \\
\hline
$U_{3}$ & $(1,1)$ $(1,4)$ $(4,1)$ $(4,4)$ & $\pm 0.8928$  & $-0.3586 \pm 0.2726i$\\
\hline
$U_{4}$ & $(1,1)$ $(1,5)$ $(5,1)$ $(5,5)$ & $\pm 0.8854$ &  $0.4610 \pm 0.0600i$ \\
\hline
$U_{5}$ &  $(1,1)$ $(1,6)$ $(6,1)$ $(6,6)$ & $\pm 0.8919$ & $0.4387 \pm 0.1098i$\\
\hline
$U_{6}$ &  $(1,1)$ $(1,7)$ $(7,1)$ $(7,7)$ &  $\pm 0.9931$ & $-0.1166 \mp 0.0145i$ \\
\hline
$U_{7}$ &  $(1,1)$ $(1,8)$ $(8,1)$ $(8,8)$ & $\pm 0.8023$ & $0.4634 \mp 0.3763i$ \\
\hline
$U_{8}$ & $(2,2)$ $(2,3)$ $(3,2)$ $(3,3)$ &  $\pm 0.3820 - 0.1754i$ & $-0.8685 \mp 0.2628i$ \\
\hline
$U_{9}$ &  $(2,2)$ $(2,4)$ $(4,2)$ $(4,4)$  &  $\pm 0.817 $ & $ -0.4289 \pm 0.3855i$ \\
\hline
$U_{10}$ &  $(2,2)$ $(2,5)$ $(5,2)$ $(5,5)$  & $\pm 0.9603 $& $-0.2731 \mp 0.0564i$ \\
\hline
$U_{11}$ &  $(2,2)$ $(2,6)$ $(6,2)$ $(6,6)$  &  $\pm 0.8358$ & $-0.5531 \pm 0.1099i$ \\
\hline
$U_{12}$ &  $(2,2)$ $(2,7)$ $(7,2)$ $(7,7)$  & $\pm 0.8158 $ & $0.5764 \pm 0.0470i$ \\
\hline
$U_{13}$ &  $(2,2)$ $(2,8)$ $(8,2)$ $(8,8)$   & $\pm 0.8756 $& $0.0043 \mp 0.4830i$\\
\hline
$U_{14}$ &  $(3,3)$ $(3,4)$ $(4,3)$ $(4,4)$  & $\mp 0.2204 - 0.0365i$& $0.9718 \pm 0.0758i$ \\
\hline
$U_{15}$ &  $(3,3)$ $(3,5)$ $(5,3)$ $(5,5)$  & $\pm 0.9334$ & $-0.2653 \mp 0.2414i$ \\
\hline
$U_{16}$ &  $(3,3)$ $(3,6)$ $(6,3)$ $(6,6)$  & $\pm 0.8020$ & $-0.5862 \pm 0.1148i$\\
\hline
$U_{17}$ &  $(3,3)$ $(3,7)$ $(7,3)$ $(7,7)$  & $\pm 0.9344$ & $0.0715 \pm 0.3489i$ \\
\hline
$U_{18}$ &  $(3,3)$ $(3,8)$ $(8,3)$ $(8,8)$ & $\pm 0.7672$ & $0.5579 \mp 0.3165i$\\
\hline
$U_{19}$ &  $(4,4)$ $(4,5)$ $(5,4)$ $(5,5)$  &  $\pm 0.0780 - 0.3382i $&$ 0.7965 \mp 0.4951i$\\
\hline
$U_{20}$ &  $(4,4)$ $(4,6)$ $(6,4)$ $(6,6)$  &  $\pm 0.9847$  & $-0.1738 \mp 0.0152i$\\
\hline
$U_{21}$ &  $(4,4)$ $(4,7)$ $(7,4)$ $(7,7)$  &   $\pm 0.6388$ &  $-0.6861 \pm 0.3482i$ \\
\hline
$U_{22}$ &  $(4,4)$ $(4,8)$ $(8,4)$ $(8,8)$  & $\pm 0.8389$  & $ 0.4026 \mp 0.3664i$ \\
\hline
$U_{23}$ &  $(5,5)$ $(5,6)$ $(6,5)$ $(6,6)$  &  $\mp 0.1500 - 0.8686i $&$ -0.3133 \pm 0.3535i$  \\
\hline
$U_{24}$ &  $(5,5)$ $(5,7)$ $(7,5)$ $(7,7)$  & $\pm 0.2687$ & $ 0.9632 \mp 0.0062i$  \\
\hline
$U_{25}$ &  $(5,5)$ $(5,8)$ $(8,5)$ $(8,8)$  &  $\pm 0.8739$  & $-0.4544 \pm  0.1725i$ \\
\hline
$U_{26}$ &  $(6,6)$ $(6,7)$ $(7,6)$ $(7,7)$  &  $\mp  0.5328 - 0.0609i $&$ -0.8269 \pm 0.1693i$\\
\hline
$U_{27}$ &  $(6,6)$ $(6,8)$ $(8,6)$ $(8,8)$  &  $\pm 0.6578$ & $ -0.7441 \mp 0.1167i$ \\
\hline
$U_{28}$ &  $(7,7)$ $(7,8)$ $(8,7)$ $(8,8)$  &  $\alpha =  0.2867 + 0.8895i$  & $\beta =   0.1921 - 0.2995i$ \\
  &     & $\gamma =    0.1800 - 0.3069i$ & $\delta = 0.9229 - 0.1473i$ \\
\hline
\end{tabular} 
\caption{Details of sub-matrices $\mathbf{u}_k$ for $\mathbf{X}_{\ell_{2,1}}$}
\label{tab:info_UL21}
\end{table}
\clearpage
\begin{table} [p]
\centering
\begin{tabular}{ |c|c|c|c|}
\hline
$U$ & $\theta$ & $\phi$ & $\lambda$ \\
\hline
$U_{1}$ & $1.57$ & $1.57$ & $2.397$ \\
\hline
$U_{2}$ & $0.002$ & $0.002$ & $-2.903$ \\
\hline
$U_{3}$ & $0.004$ & $0.004$ & $3.047$ \\
\hline
$U_{4}$ & $0.004$ & $0.004$ & $-2.137$ \\
\hline
$U_{5}$ & $-$ & $-$ & $-$ \\
\hline
$U_{6}$ & $-$ & $-$ & $-$ \\
\hline
$U_{7}$ & $-$ & $-$ & $-$ \\
\hline
$U_{8}$ & $1.571$ & $1.571$ & $-3.044$ \\
\hline
$U_{9}$ & $0.004$ & $0.004$ & $0.184$ \\
\hline
$U_{10}$ & $0$ & $0$ & $-1.856$ \\
\hline
$U_{11}$ & $-$ & $-$ & $-$ \\
\hline
$U_{12}$ & $0.006$ & $0.006$ & $-2.646$ \\
\hline
$U_{13}$ & $-$ & $-$ & $-$ \\
\hline
$U_{14}$ & $0.004$ & $0.004$ & $0.881$ \\
\hline
$U_{15}$ & $0.419$ & $0.419$ & $1.767$ \\
\hline
$U_{16}$ & $0.036$ & $0.036$ & $-2.289$ \\
\hline
$U_{17}$ & $0$ & $0$ & $-2.655$ \\
\hline
$U_{18}$ & $0.002$ & $0.002$ & $-2.865$ \\
\hline
$U_{19}$ & $1.504$ & $1.504$ & $2.109$ \\
\hline
$U_{20}$ & $1.497$ & $1.497$ & $0.388$ \\
\hline
$U_{21}$ & $0.453$ & $0.453$ & $1.387$ \\
\hline
$U_{22}$ & $0.005$ & $0.005$ & $0.285$ \\
\hline
$U_{23}$ & $1.496$ & $1.496$ & $0.969$ \\
\hline
$U_{24}$ & $0.841$ & $0.841$ & $0.513$ \\
\hline
$U_{25}$ & $1.563$ & $1.563$ & $-3.003$ \\
\hline
$U_{26}$ & $1.567$ & $1.567$ & $-0.585$ \\
\hline
$U_{27}$ & $0.802$ & $0.802$ & $2.859$ \\
\hline
$U_{28}$ & $0.011$ & $0.011$ & $-1.747$ \\
\hline
\end{tabular}
\caption{Angles of gates for $\ell_{1}$-norm.}
\label{tab:angle_UL1}
\end{table}
\begin{table} [p]
\centering
\begin{tabular}{ |c|c|c|c|  } 
\hline
$U$ & $\theta$ & $\phi$ & $\lambda$ \\
\hline
$U_{1}$ & $0.571$ & $0.571$ & $-0.357$ \\
\hline
$U_{2}$ & $0.577$ & $0.577$ & $-3.026$ \\
\hline
$U_{3}$ & $0.467$ & $0.467$ & $-0.65$ \\
\hline
$U_{4}$ & $0.483$ & $0.483$ & $-3.012$ \\
\hline
$U_{5}$ & $0.469$ & $0.469$ & $-2.896$ \\
\hline
$U_{6}$ & $0.118$ & $0.118$ & $0.123$ \\
\hline
$U_{7}$ & $0.64$ & $0.64$ & $2.459$ \\
\hline
$U_{8}$ & $1.137$ & $1.137$ & $-0.136$ \\
\hline
$U_{9}$ & $0.615$ & $0.615$ & $-0.732$ \\
\hline
$U_{10}$ & $0.283$ & $0.283$ & $0.203$ \\
\hline
$U_{11}$ & $0.599$ & $ 0.599$ & $-0.196$ \\
\hline
$U_{12}$ & $0.617$ & $0.617$ & $-3.06$ \\
\hline
$U_{13}$ & $0.504$ & $0.504$ & $1.58$ \\
\hline
$U_{14}$ & $1.346$ & $1.346 $ & $0.242$ \\
\hline
$U_{15}$ & $0.367$ & $0.367$ & $0.738$ \\
\hline
$U_{16}$ & $0.64$ & $0.64$ & $-0.193$ \\
\hline
$U_{17}$ & $0.364$ & $0.364 $ & $-1.773$ \\
\hline
$U_{18}$ & $0.696$ & $0.696$ & $2.626$ \\
\hline
$U_{19}$ & $1.216$ & $1.216$ & $1.241$ \\
\hline
$U_{20}$ & $0.175$ & $0.175$ & $0.087$ \\
\hline
$U_{21}$ & $0.878$ & $0.878$ & $-0.47$ \\
\hline
$U_{22}$ & $0.576$ & $0.576$ & $2.403$ \\
\hline
$U_{23}$ & $0.492$ & $0.492$ & $-2.587$ \\
\hline
$U_{24}$ & $1.299$ & $1.299$ & $3.135$ \\
\hline
$U_{25}$ & $0.508$ & $0.508$ & $-0.363$ \\
\hline
$U_{26}$ & $1.005$ & $1.005$ & $-0.316$ \\
\hline
$U_{27}$ & $0.853$ & $0.853$ & $0.156$ \\
\hline
$U_{28}$ & $0.364$ & $0.364$ & $-2.883$ \\
\hline
\end{tabular}
\caption{Angles of gates for $\ell_{2,1}$-norm.}
\label{tab:angle_UL21}
\end{table}

\section{ Conclusion} \label{sec:conclusion}

In conclusion, we introduced a new approach for obtaining transformation matrices that transform a given initial state to a desired final state. Through the formulation of a non-convex objective function and the utilization of the ADMM algorithm, we successfully address the challenge of incorporating sparsity, unitary constraints, and row/column-wise unit sum constraints in the optimization process. Our algorithm not only achieves sparsity in the transformation matrices but also preserves essential properties, distinguishing it from previous techniques.

Through numerical simulations, we validate the effectiveness of our proposed approach in solving sparse transformation optimization problems while satisfying the required constraints. We demonstrate the use of both the individual sparsity $\ell_1$ norm and the row sparsity $\ell_{2,1}$ norm in searching for a sparse transformation matrix. Both sparsity norms effectively promote sparsity in the context of quantum state transformation, each with its own advantages and limitations.

Furthermore, this research highlights the potential of sparsity and group sparsity optimization techniques in enhancing quantum state transformations. By controlling the trade-off between sparsity and the preservation of essential properties, we can achieve the desired approximation for various applications.

In summary, our approach offers a new and effective solution for obtaining transformation matrices, enabling efficient quantum state transformations while preserving essential properties. We anticipate that our work will stimulate ongoing progress and future investigations in combining sparse optimization and quantum computing to enhance the efficiency and practicality of quantum information processing.

\section*{Acknowledgments}

This work is supported by the National Natural Science Foundation of China (Grant No. 11874312), the Research Grants Council of Hong Kong (CityU 11304920), the Guangdong Provincial Quantum Science Strategic Initiative (Grant No. GDZX2203001, GDZX2303007), and the Innovation Program for Quantum Science and Technology (Grant No. 2021ZD0302300).

\appendix
\section{Comparison of sparsity for original (unnormalized) vector}\label{appx:appxA}

\begin{table}
\centering
    \begin{tabular}{|c|c|c|c|c|c|c|}
    \hline
    \multicolumn{7}{|c|}{${\ell_{1}}$-Norm}\\
    \hline
    \textbf{Number of Qubits} &  n = 2 & n = 3 &  n = 4 &  n = 5 &  n = 6 &  n = 7 \\
    \hline
    {Converged value of $L$} & $1.513$ & $2.91$ & $6.537$ & $13.403$ & $25.446$ & $48.133$ \\
    \hline
    {No. of iterations required for convergence} & $>1300$ & $>1600$ & $>1100$ & $>1100$ & $>1200$ & $>1000$ \\
    \hline
    {Percentage of Non-Zero Elements} & $<76 \%$ & $<74 \%$ & $<70 \%$ & $<65 \%$ & $<60 \%$ & $<45 \%$  \\
    \hline
    \end{tabular}
    
    \vspace{10pt}

    \begin{tabular}{|c|c|c|c|c|c|c|}
    \hline
    \multicolumn{7}{|c|}{${\ell_{2,1}}$-Norm}\\
    \hline
    \textbf{Number of Qubits} &  n = 2 & n = 3 &  n = 4 &  n = 5 &  n = 6 &  n = 7 \\
    \hline
    {Converged value of $L$} & $0.972$ & $1.706$ & $3.873$ & $9.399$ & $19.933$ & $42.719$ \\
    \hline
    {No. of iterations required for convergence} & $>100$ & $>100$ & $>200$ & $>600$ & $>1200$ & $>2000$ \\
    \hline
    {Number of Non-Zero Elements} & $16$ & $64$ & $256$ & $1024$ & $4096$ & $16384$  \\
    \hline
    \end{tabular}
\caption{Comparison of ${\ell_1}$ and $\ell_{2,1}$ norm on random unnormalized vectors $\mathbf{A}$ and $\mathbf{C}$.}
\label{tab:un_compareL}
\end{table}

The purpose of this appendix is to demonstrate the performance and convergence of our ADMM algorithm on Eq.~\eqref{eq:final} with unnormalized vectors. The results are shown in Table~\ref{tab:un_compareL}. In this case, the values of elements in matrix $\mathbf{A}$ and vector $\mathbf{C}$ have larger absolute magnitudes compared to the penalty terms. We aim to investigate the impact of unnormalized vectors on the same problem and compare the performance of the $\ell_{1}$ norm and $\ell_{2,1}$ norm penalties.

Based on this experiment, it is evident that the $\ell_{2,1}$ norm consistently outperforms the $\ell_{1}$ norm penalty. It achieves better optimal solutions while satisfying the given constraints, suggesting that the $\ell_{2,1}$ norm is a superior choice for introducing sparsity in these problems. However, similar to the experiment on normalized vectors, the $\ell_{1}$ norm proves to be effective in producing a more sparse result. Therefore, depending on the specific objectives of the problem, both penalty methods have their strengths and can be employed accordingly.

\newpage
\section{Comparison of sparsity for least square objective problem}\label{appx:appxB}

\begin{table}[htbp]
    \centering
    \begin{tabular}{|c|c|c|c|c|}
    \hline
   \textbf{Matrix Size $(m \times n)$} & $10 \times 5$ &   $5 \times 10$ & $100 \times 50$ & $50 \times 100$    \\
        \hline
        {Converged value of $L$} & $85.679$ & $91.969$ & $9036.237$ & $9935.633$
 \\
        \hline
        {No. of iterations required for convergence}  & $>600$ & $>900$ & $>9500$ & $<100$\\
        \hline
        {Percentage of Non-zero Elements}  & $<64\%$ & $<27\%$ & $<32\%$ & $<11 \%$\\
        \hline
    \end{tabular}
\caption{Comparison of ${\ell_1}$ least square objective problem with unitary constraint with different sizes.}
\label{tab: LS1_compareL}
\end{table}

\begin{table}[htbp]
    \centering
    \begin{tabular}{|c|c|c|c|c|}
    \hline
   \textbf{Matrix Size $(m \times n)$} & $10 \times 5$ &   $5 \times 10$ & $100 \times 50$ & $50 \times 100$    \\
        \hline
        {Converged value of $L$} &$60.607$ &$ 25.353$  & $4582.105$ & $1355.355$ \\
        \hline
        {No. of iterations required for convergence}  & $>400$ & $>500$ & $>3200$ & $>4000$\\
        \hline
        {Number of Non-Zero Elements}  & $25$ & $100$ & $2500$ & $10000$\\
        \hline
    \end{tabular}
\caption{Comparison of ${\ell_{2,1}}$ least square objective problem with unitary constraint with different sizes.}
\label{tab: LS2_compareL}
\end{table}

In this appendix, we show the performance and convergence of the proposed algorithm in other objective problems, highlighting its effectiveness beyond our specific application. In this context, We aim to simulate the recovery of matrix $\mathbf{X}$ from a given set of measurements represented by matrix $\mathbf{B}$ and sensing operator $\mathbf{A}$. The objective function we consider is $\|\mathbf{A}\mathbf{X}-\textbf{B}\|^2$, subject to unitary and row/column-wise unit sum constraints on $\mathbf{X}$. This problem formulation has applications in unitary matrix sensing, which is relevant in fields such as quantum information science, signal processing, and system identification.

We compare the effectiveness of the $\ell_1$ and $\ell_{2,1}$ sparsity norms in this problem. Table~\ref{tab: LS1_compareL} and~\ref{tab: LS2_compareL} present a comparison of these sparsity norms for different sizes and shapes of matrices.
For the size comparison, we consider small-sized matrices with dimensions $5\times 10$ and $10\times 10$, as well as larger matrices with dimensions $50\times 100$ and $100\times 100$. This allows us to evaluate the performance of the algorithm across different matrix sizes.
Additionally, We consider both ``tall matrix'' and ``wide matrix'' shapes, providing further insights into how the algorithm behaves with varying aspect ratios.

\section{Gate Efficiency in Quantum Circuit Decomposition}\label{appx:appxD}

\begin{table}
\centering
\begin{tabular}{ |c|c|c|c|c|c|  }
 \hline
Number of qubits & $3$ & $4$ & $5$ & $6$ & $7$ \\
 \hline
Maximum Number of Gates & $28$ & $120$ & $630$ & $3960$ & $28560$ \\ 
 \hline
Average Number of Gates & $<26$ & $<110$ & $<500$ & $<2000$ & $<8000$ \\
 \hline
\end{tabular}
\caption{Average number of decomposed gates for unitary matrices.}
\label{tab:max_gate}
\end{table}

The purpose of this appendix is to demonstrate the impact of sparsity on the decomposition process of unitary matrices in quantum circuits. It presents an analysis of the average number of gates required to construct the transformation matrix $\mathbf{X}$ with $\ell_1$ norm using our proposed ADMM algorithm for different numbers of qubits. 
Table~\ref{tab:max_gate} showcases the average number of gates required for constructing the transformation matrix $\mathbf{X}$ compared to the maximum number of gates required when considering the same number of qubits.

From Table~\ref{tab:max_gate}, it shows that while the number of gates required for constructing the unitary matrix $\mathbf{X}$ decreases significantly compared to the maximum number required, there is still a substantial increase in gate requirements with each additional qubit. This observation highlights the growing complexity associated with decomposing unitary matrices as the number of qubits increases. It emphasizes the necessity for optimizing gate usage and developing efficient decomposition techniques to minimize the resource overhead in quantum circuits.

\end{document}